\documentclass[prl,nofootinbib,notitlepage,superscriptaddress]{revtex4-1}
\usepackage{amsmath}
\usepackage{braket}
\usepackage{graphicx,color}
\usepackage{tikz}

\newcommand{\arrrho}[2]{\begin{tikzpicture} 
  \draw[to-to reversed,line width=1pt] (-1,0)  -- (0,0); 
  \node [left] at (-.9,0) {${}_{#1}$};
  \node [right] at (-.1,0) {${}_{#2}$};
\end{tikzpicture}}
\newcommand{\arrtau}[2]{\begin{tikzpicture} 
  \draw[to reversed-to reversed,line width=1pt] (-1,0)  -- (0,0); 
  \node [left] at (-.9,0) {${}_{#1}$};
  \node [right] at (-.1,0) {${}_{#2}$};
\end{tikzpicture}}
\newcommand{\arrtaus}[2]{\begin{tikzpicture} 
  \draw[to-to,line width=1pt] (-1,0)  -- (0,0); 
  \node [left] at (-.9,0) {${}_{#1}$};
  \node [right] at (-.1,0) {${}_{#2}$};
\end{tikzpicture}}

\newcommand{\arrrhoend}[2]{\begin{tikzpicture} 
  \draw[to-to reversed,line width=1pt] (-1,0)  -- (0,0); 
  \draw [red, ultra thick] (.1,0) circle [radius=0.3];
  \node [left] at (-.9,0) {${}_{#1}$};
  \node [right] at (-.1,0) {${}_{#2}$};
\end{tikzpicture}}
\newcommand{\arrrhobeg}[2]{\begin{tikzpicture} 
  \draw[to-to reversed,line width=1pt] (-1,0)  -- (0,0); 
  \draw [red, ultra thick] (-1.1,0) circle [radius=0.3];
  \node [left] at (-.9,0) {${}_{#1}$};
  \node [right] at (-.1,0) {${}_{#2}$};
\end{tikzpicture}}

\begin{document}

%\preprint{}

\title{Semiclassical Approach to Dynamics of Interacting Fermions}


\author{Shainen M. Davidson}
\email[Corresponding author: ] {shainen@bu.edu}
\affiliation{Department of Physics, Boston University, 590 Commonwealth Ave., Boston, MA 02215, USA}

\author{Dries Sels}
\affiliation{Department of Physics, Boston University, 590 Commonwealth Ave., Boston, MA 02215, USA}
\affiliation{TQC, Universiteit Antwerpen, Universitetisplein 1, B-2610 Antwerpen, Belgium}

\author{Anatoli Polkovnikov}
\affiliation{Department of Physics, Boston University, 590 Commonwealth Ave., Boston, MA 02215, USA}


\date{\today}

\begin{abstract}
Understanding the behavior of interacting fermions is of fundamental interest in many fields ranging from condensed matter to high energy physics. Developing numerically efficient and accurate simulation methods is an indispensable part of this. Already in equilibrium, fermions are notoriously hard to handle due to the sign problem. Out of equilibrium, an important outstanding problem is the efficient numerical simulation of the dynamics of these systems. In this work we develop a new semiclassical phase-space approach (a.k.a. the truncated Wigner approximation) for simulating the dynamics of interacting fermions in arbitrary dimensions. As fermions are essentially non-classical objects, a phase-space is constructed out of all fermionic bilinears. Classical phase-space is thus comprised of highly non-local (hidden) variables representing these bilinears, and the cost of the method is that it scales quadratic rather than linear with system size. We demonstrate the strength of the method by comparing the results to the exact quantum dynamics of fermion expansion in the Hubbard model and quantum thermalization in the Sachdev-Ye-Kitaev (SYK) model for small systems, where the semiclassics nearly perfectly reproduces correct results. We furthermore analyze fermion expansion in a larger, intractable by exact methods, 2D Hubbard model, which is directly relevant to recent cold atom experiments.
\end{abstract}

%\keywords{}

\maketitle

\paragraph*{Introduction}
The study of non-equilibrium quantum phenomena is at the forefront of physics. This research is driven by the increasing ability to isolate quantum systems from their environment, resulting in nearly perfect unitary dynamics. Although much progress has been made in various directions, our understanding is limited by the theoretical tools available to simulate large quantum systems. Exact diagonalization is limited to really small system sizes. For larger systems one typically relies on density-matrix renormalization group (DMRG), dynamical mean-field theory (DMFT), density functional theory (DFT) or semi-classical phase space methods. Unfortunately, DMRG type methods are only efficient in one dimension and sufficiently close to the ground states~\cite{Schollwock2004}. DMFT methods have been recently extended to non-equilibrium systems, but the results usually work only qualitatively~\cite{Hideo2014}.
 Phase-space methods provide an efficient and accurate tool to study quantum dynamics near the classical limit or at short times. They have found many applications in quantum optics, atomic physics, quantum chemistry, high-energy physics, and other areas, see e.g. Refs.~\cite{Hillery1984, Coker_98, Blakie2008, Polkovnikov2010a, Berges2014}. Recently there have been interesting generalizations of these methods to tackle the dynamics of spins using discrete phase space representations~\cite{Schachenmayer2015} or combining them with Lindblad dynamics for open systems~\cite{Rey2014}. These methods generally work well either near the semiclassical limit where quantum fluctuations can be treated perturbatively or near non-interacting limits where the Heisenberg equations of motion for operators representing degrees of freedom are nearly linear and the commutators can be replaced by Poisson brackets. However there is no obvious extension of these methods to fermions because they do not have a natural classical representation. Moreover it is clear that any local representation of fermions is generally incompatible with local classical dynamics because the wave function of e.g, two fermions, taken around each other, acquires a negative sign even if the fermions never directly interact. This sign crucially affects fermion properties, such as leading to the Pauli exclusion principle.

In this work we develop an efficient semiclassical phase-space approach for fermions using string variables, resulting in a simple and tractable method for simulating their dynamics. While fermionic phase-space distributions were already introduced in the context of P- and Q-representations~\cite{Corney2006,Rahav2009,Rosales2015}, the Wigner function for fermions has only been introduced as a function of Grassmann variables~\cite{Cahill1999,Stanislaw2013}. Anticommuting Grassmann variables only allow for manipulations on a formal level, seriously hampering any practical use beyond perturbation theory. Here we focus on the Wigner-Weyl quantization of fermions and present a Grassmann variable free formulation of the problem. We illustrate the method by simulating expansion of interacting fermions in a lattice both in one and two dimensions and show very good agreement with exact diagonalization results in small systems. Moreover, we show that our method essentially exactly describes thermalization of the Sachdev-Ye-Kitaev (SYK) model. This suggests that this model provides a natural realization of the non-linear classical Hamiltonian dynamics of interacting strings.

Let us emphasize that since the string variables are fluctuating what we are discussing goes beyond any mean-field approximation, where these variables are simply replaced by their expectation values. As with any other semiclassical method it is expected that its accuracy is relying on existence of slow e.g. hydrodynamic degrees of freedom. While our intuition about the dynamics of the strings is limited and we cannot present a formal estimate of the error at the moment, we demonstrate, by comparing directly to exact dynamics, that this intuition is correct and the method works very well in various experimentally relevant interacting models. We also apply the method to two-dimensional systems and show that its predictions are consistent with recent experiments. We are not aware of any other existing competing methods which can lead to a comparable accuracy.

\paragraph*{f-TWA}
The inherent probablistic nature of observations in quantum theory has sparked many attempts at reformulating the principles of quantum theory in purely statistical terms. It is indeed tempting to try to replace quantum fluctuations with statistical fluctuations. However, any representation of quantum theory as a statistical phase-space theory deals with the fundamental problem that it must represent the joint probability of measuring non-commuting observables. While this is possible, it makes phase-space distributions not unique for a given state as one must specify additional rules in how to deal with these non-commuting observables. Essentially one has to specifiy the order in which one is going to measure the observables. Here we will consider symmetric ordering, the so called Wigner-Weyl representation. 

For canonical variables, the Wigner-Weyl correspondence provides a unique map between any function of operators and a function over classical phase-space, where for each degree of freedom there is a pair of conjugate phase-space variables $(x,p)$~\cite{Polkovnikov2010a,Hillery1984}. Expectation values can be calculated by classically averaging the Weyl symbol of any operator with the Wigner function (the Weyl symbol of the density matrix):
\begin{gather}
\langle \hat
O\rangle = \int d\vec x d\vec p \:
\mathcal{W}(\vec x ,\vec p)
\mathcal{O}(\vec x,\vec p),   
\end{gather}
where $\mathcal{W}(\vec x,\vec p)$ is the Wigner function and $\mathcal{O}(\vec x,\vec p)$ is the Weyl symbol of $\hat O$. Any quantum dynamical problem corresponds to a problem of propagating the Wigner function in time. In general, time evolution of the Wigner function is hard to handle as the action for its propagator contains highly non-linear terms~\cite{Sels2013}. However, near the classical limit one can use the so-called truncated Wigner approximation
(TWA)~\cite{Blakie2008,Polkovnikov2010a}, where the Wigner
function, like a classical probability distribution, is conserved on
classical trajectories determined by the Poisson bracket:
\begin{align}
  \label{eq:cl}
  \frac{d x_i}{d t}= \left\lbrace x_i, \mathcal{H} \right\rbrace_P, \quad   \frac{d p_i}{d t}= \left\lbrace p_i, \mathcal{H} \right\rbrace_P,
\end{align}
where $\mathcal{H}$ is the Weyl (symmetrically) ordered Hamiltonian. Note that the phase space variables $(x_i, p_i)$ are not simply the expectation values of the associated operators, but are semi-classical variables. Even though they dynamically never deviate from the classical trajectory, quantum correlations between operators are still encoded in the initial Wigner function. This description applies to any set of canonical variables, not just momenta and coordinates of particles. It can therefore be used to describe any bosonic system and it has found important applications in quantum optics and ulta-cold atoms (see Methods). 

In the case of fermions, however, the anti-commuting nature of the fermions does not allow a naive phase-space formulation in terms of cannonical fermionic phase-space variables for each fermionic mode. Unlike bosons, fermionic operators cannot be represented by simple complex numbers, since any product of two fermion operators picks up a minus sign upon exchanging the order of the operators.  However, as a consequence of the same anti-commutation relations, any physical observable must always contain an even number of fermionic operators. Hence, the simplest possible observable fermionic operators are bilinears of creation and anihilation operators and we treat those as the classical phase space variables. In that respect we follow Heisenberg's wise words: \emph{the statistical predictions of quantum theory are thus significant only when combined with experiments which are capable of observing the phenomena treated by the statistics.} 

 Consider a system of $N$ fermionic modes, described by the creation (annihilation) operators $\hat c_\alpha^\dagger$ ($\hat c_\alpha$), where $\alpha$ labels all single-particle quantum numbers associated with the fermions,  such as spin, lattice index, or single-particle orbital in atoms or molecules. Classical phase-space is now made up of $N^2$ variables encoded in the matrices $\rho$ and $\tau$:
\begin{equation}
\frac{1}{2}\left(\hat c_\alpha^\dagger \hat c_\beta-\hat c_\beta \hat c_\alpha^\dagger \right) \mapsto \rho_{\alpha\beta}, \quad \hat c_{\alpha} \hat c_{\beta}  \mapsto \tau_{\alpha\beta},\quad \hat c_{\beta}^\dagger \hat c_{\alpha}^\dagger  \mapsto \tau_{\alpha\beta}^\ast.
\end{equation}
Note that these variables are maximally non-local: the price we pay for representing fermions in a classical phase-space. Physically, the expectation values of the operators $\rho$ and $\tau$ represent two-point correlators. Their classical equations of motion again are determined by the Poisson bracket for fermion variables:
\begin{gather}
  \frac{d}{dt}\rho_{\alpha\beta} = \left\lbrace \rho_{\alpha\beta}, \mathcal{H} \right\rbrace_{fP}, \quad   \frac{d}{dt}\tau_{\alpha\beta}= \left\lbrace \tau_{\alpha\beta}, \mathcal{H} \right\rbrace_{fP},
\label{eq:TWAeqmot}
\end{gather}
which can be derived from \eqref{eq:cl} (see Methods). The easiest way to visualize the Poisson brackets is graphical: the fermionic phase-space variables can be represented as strings, with arrows at the ends pointing out (in) to represent creation (anihilation) operators:
\begin{align}
  \begin{array}{c}\rho_{\alpha\beta}\\\arrrho{\alpha}{\beta}\end{array} &&
  \begin{array}{c}\tau_{\alpha\beta}\\\arrtau{\alpha}{\beta}\end{array} &&
  \begin{array}{c}\tau^*_{\alpha\beta}\\\arrtaus{\alpha}{\beta}\end{array}
\end{align}
The Poisson bracket is then the result of connecting all the ends which terminate at the same point and fit together. For example, for a local spin interaction at a site $i$ with form $\rho_{i\uparrow i\uparrow}\rho_{i\downarrow i\downarrow}$, we would generate dynamics for the variable $\rho_{j\uparrow i\uparrow}$ (where $j \neq i$),
\begin{multline}
  \frac{d}{dt}\rho_{j\uparrow i\uparrow} = \left\lbrace \rho_{j\uparrow i\uparrow}, \rho_{i\uparrow i\uparrow}\rho_{i\downarrow i\downarrow}\right\rbrace_{fP} 
= \left\lbrace \arrrhoend{j\uparrow}{i\uparrow}, \begin{array}{c}\arrrhobeg{i\uparrow }{i\uparrow}\\\arrrho{i\downarrow }{i\downarrow}\end{array}\right\rbrace_{fP} = \begin{array}{c}\arrrho{j\uparrow }{i\uparrow}\\\arrrho{i\downarrow }{i\downarrow}\end{array} = \rho_{j\uparrow i\uparrow}\rho_{i\downarrow i\downarrow}.
\label{eq:poisson_fermion}
\end{multline}
One can also visualize these rules by noticing that two outgoing strings (left) are replaced by a new string (right) such that three of them form a directed triangle. In the dynamics generated by the Hamiltonian, we see locality manifest: while the phase space variables are not local, the fermionic Poison bracket of the variables are only affected by terms in the Hamiltonian which share a common index.

The precise form of the Wigner function is not feasble to reproduce, and generally contains negative values that complicate numerics; however, within the accuracy of TWA it is sufficient to use an approximate Wigner function to calculate expectation values:
 \begin{align}
\langle \hat
O(t)\rangle\approx\int d \rho d \tau
\mathcal{W}(\rho,\tau)
\mathcal{O}(\rho_{cl}(t),\tau_{cl}(t)).
\label{eq:TWAexp} 
 \end{align}
We use a gaussian distribution for $\mathcal{W}(\rho,\tau)$, with mean and variance fixed by initial conditions. Note that, on the Gaussian level, the Wigner function always factorizes: $\mathcal{W}(\rho,\tau)=\mathcal{W}_\rho(\rho)\mathcal{W}_\tau(\tau)$. If the initial distribution corresponds to the equilibrium Fermi-sea then the noise on $\rho$ can be understood as originating from particle hole excitations. Whenever the Fermi-sea is either full or empty there is no room for these excitations so the covariance of $\rho$ vanishes. In contrast there is always noise on $\tau$ as it represents two-particle fluctuations, for which there is room in any state (see Methods for details).
\paragraph*{Results}

As a test of the method, we look at two qualitatively different phenomena: fermion expansion in a lattice and thermalization of the SYK model. A third example, of a two-channel model containing coupled bosonic and fermionic degrees of freedom, is presented in the supplementary information. 

First we examine expansion dynamics of fermions in the Hubbard model with long-range and nearest-neighbour hopping, in both 1 and 2 dimensions. Such a system has been experimentally implemented in, e.g., \cite{Schneider2012a}, and we know of no other effective numerical methods for the 2d case.

The dynamics are governed by the Hamiltonian
\begin{gather}
  \hat H_\text{Hub} = - \sum_{ij\sigma} J_{ij} \left(\hat c^\dagger_{i\sigma} \hat c_{j\sigma} + \text{h.c} \right) + U \sum_i \hat n_{i\uparrow} \hat n_{i\downarrow},
\end{gather}
where
\begin{gather}
  J_{ij} = \frac{t}{|i-j|^\alpha}.
\end{gather}
For $\alpha \rightarrow \infty$, we recover the original Hubbard model with nearest-neighbour interactions. 

When we convert this Hamiltonian to a classical one in terms of fTWA variables, there is ambiguity in how we write the interation term: if we use the full set of bilinear variables, then we can decompose the interaction into both ``superconducting'' variables $\tau$ and ``density'' variables $\rho$. However, the $\rho$ sector is a subgroup of the complete group, so we can choose to only use these variables for the problem we are considering, which is not one in which we expect superconducting considerations to play a role.

Thus the classical (fTWA) Hamiltonian takes the form
\begin{gather}
\mathcal{H}_\text{Hub}= - \sum_{ij\sigma} J_{ij} \left(\rho_{i\sigma,j\sigma} + \text{h.c} \right) + U \sum_i \left(\rho_{i\uparrow,i\uparrow}+1/2\right)\left(\rho_{i\downarrow,i\downarrow}+1/2\right).
\end{gather}

To compare to exact dynamics, we look at a 16 site system at 1/4 filling in 1d with open boundaries. We initialize the system with the four center sites fully occupied. We then quench instantly to the final hopping and interaction strength. In all cases the fTWA will be exact for the non-interacting case ($U=0$).

In Fig.~\ref{fig:1d}  we compare exact dynamics and fTWA dynamics with nearest-neighbour and long-range coupling. The fTWA remains qualitatively accurate at late times, even with a significant interaction strength of up to $U=3$. As expected, the results are slightly better with long-range interactions, since as the connectivity of the system increases, we get closer to the mean-field, and hense semi-classical, limit. For the same reason, we expect results to be more accurate in 2 dimensions instead of 1.

\begin{figure}[h]
  \includegraphics[width=\linewidth]{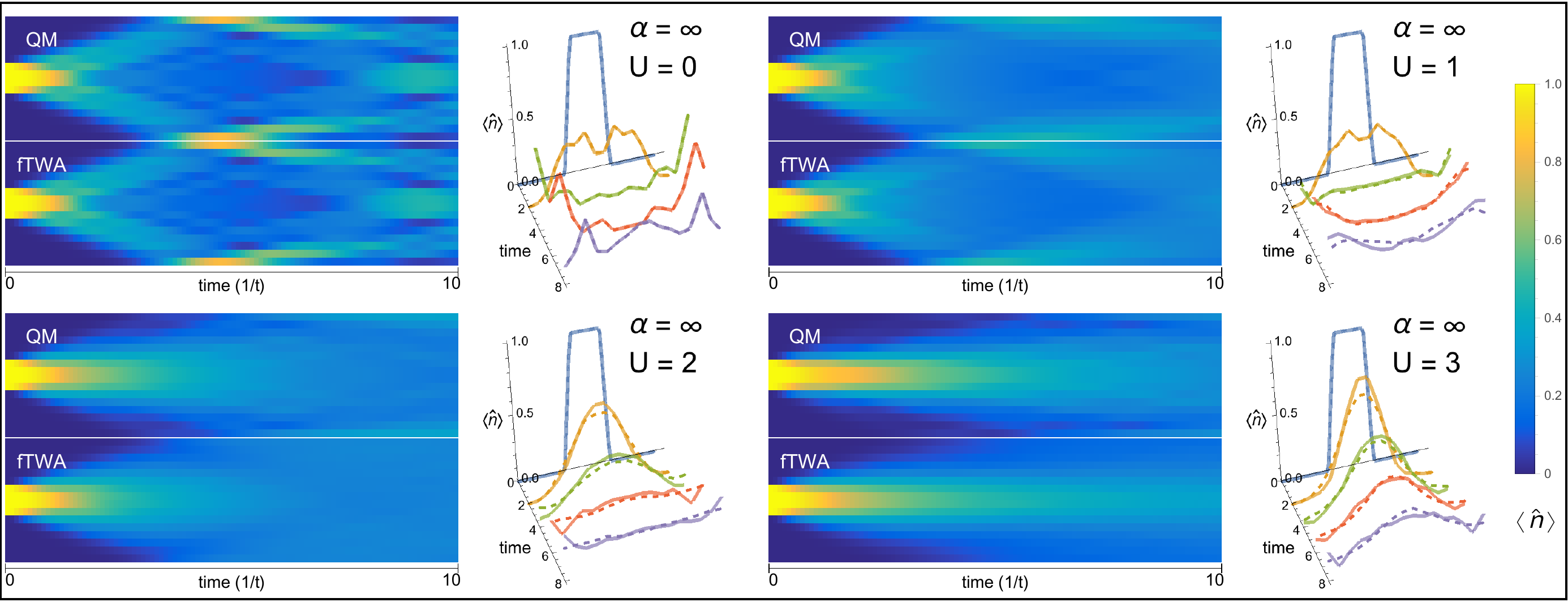}

  \includegraphics[width=\linewidth]{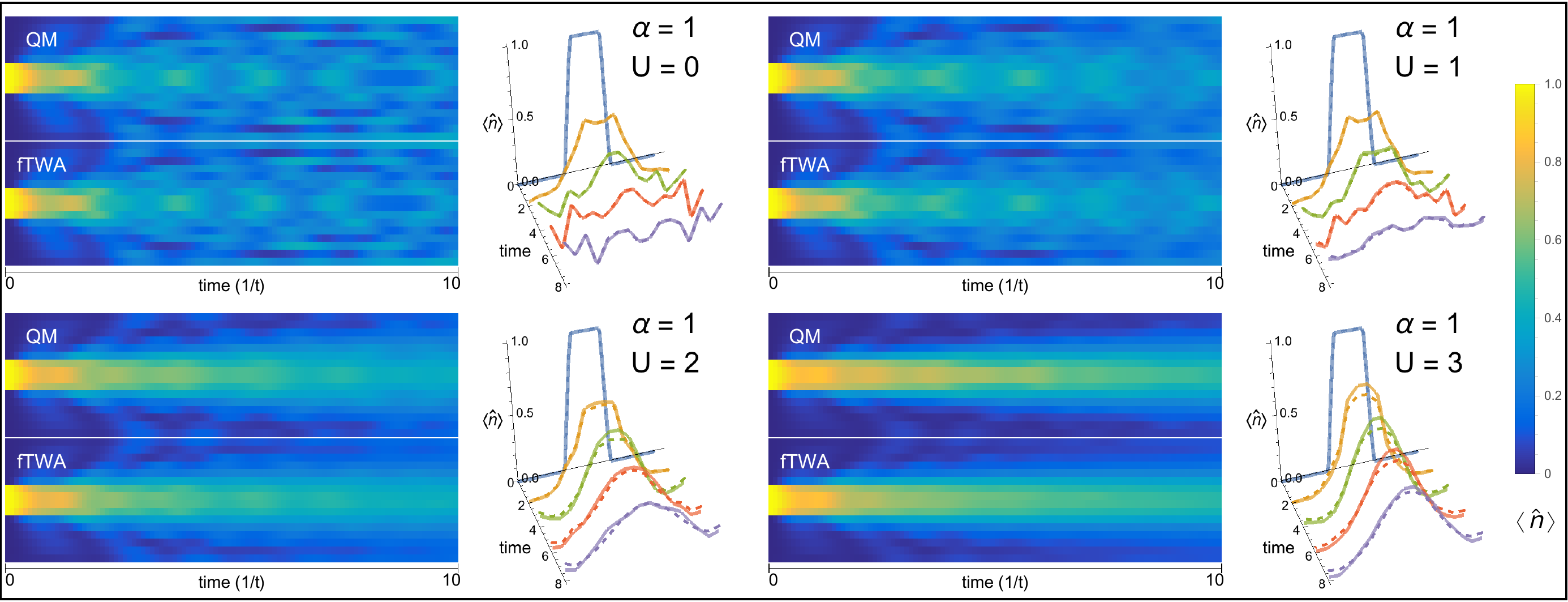}
  \caption{Comparing the fermion occupation dynamics generated by fTWA with the exact case. Top is for nearest-neighbour coupling and bottom is for long-range hopping with $\alpha =1$. For detailed comparison, profiles are shown at times $t/J = 0,2,4,8$; exact is solid and fTWA is dashed. As the interaction $U$ is increased, the error in the fTWA result grows, while the case with no interation has no error.}
  \label{fig:1d}
\end{figure}

In Fig.~\ref{fig:2d} we demonstrate fTWA results for a $12 \times 12$ 2d-system, where exact methods do not apply. In this case we show density plots at $t/J = 1.5~(t/J = 6)$ for nearest-neighbour (long-range $\alpha=1$) hopping.  As in \cite{Schneider2012a}, when there are no interactions the fermions adopt the square symmetry of the lattice, while the interaction encourages the fermion cloud to adopt a circular profile. Just like in the experiment, we observe a slowing down of the expansion with increasing interactions. In order to produce Fig.~\ref{fig:2d}, we numerically integrated Eq.~\eqref{eq:TWAeqmot} for 1000 Monte Carlo samples of the initial Gaussian Wigner distribution. Propagating one of those samples amounts to solving a set of 41472 coupled non-linear differential equations. 

\begin{figure}[h]
  \centering
  \includegraphics[width=\linewidth]{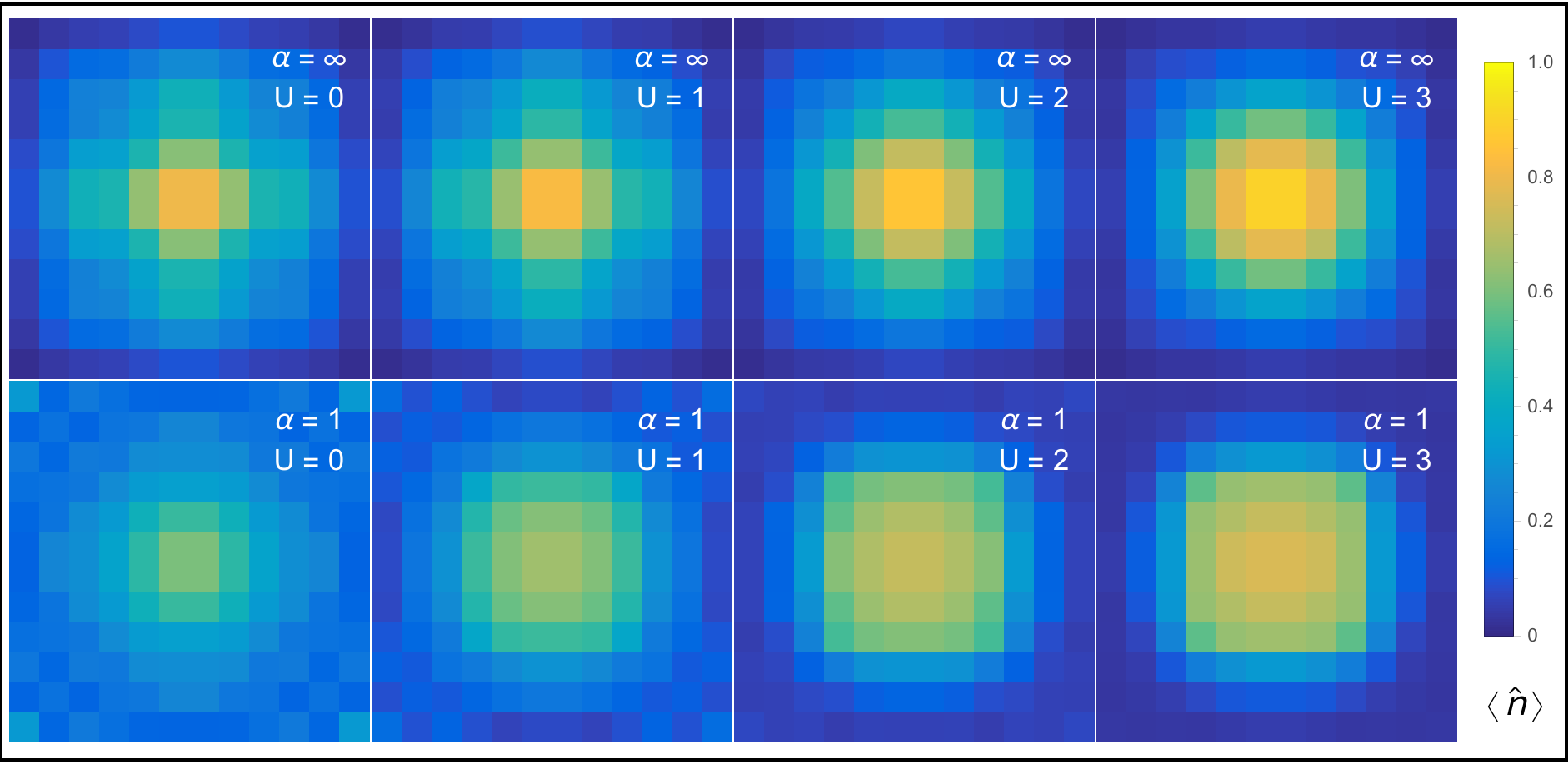}
  \caption{Denstity plots of fTWA data for a $12\times 12$ system, with nearest-neighbour and long-range couplings at different interaction strengths. The nearest-neighbour (long-range $\alpha=1$) results are time slices at $t/J = 1.5 ~(t/J = 6)$.}
  \label{fig:2d}
\end{figure}

As another nontrivial application of fTWA we consider the Sachdev-Ye-Kitaev (SYK) model, which has recently become a paradigm for studying so called bounds on chaos~\cite{Sachdev2015}:
\begin{gather}
  \hat H_{SYK} = \frac{1}{(2N)^{3/2}}\sum_{ijkl} J_{ij;kl} \hat c^\dagger_i\hat c^\dagger_j \hat c_k\hat c_l.
\end{gather}
Here $J$ are complex, independent Gaussian random numbers with zero mean, satisfying the following constraints
\begin{gather}
  J_{ij;kl} = -J_{ij;lk} = -J_{ji;kl}\\
  J_{ij;kl} = J_{kl;ij}^*,\\
 \overline{|J_{ij;kl}|^2} = J^2.
\end{gather}
The SYK model has also recently attracted tremendous attention as a tractable example of a strange metal which allows for a holographic dual description. While exactly solvable in the large-N limit, the system cannot be described in terms of quasi-particles. It is therefore interesting to check whether fTWA can capture the dynamics of the model. We found empirically that fTWA is accurate if we represent the interaction term in the Hamiltonian with superconducting $\tau$ variables. Because the latter do not for a subgroup, unlike in the previous example we have to use all fermionic string variables. In general the choice of optimal representation of the interaction term through the $\tau$ and $\rho$ variables remains an open question  (see Methods for more details). To examine the thermalizing dynamics, we fill some fraction of sites in a product state, and then measure the average occupancy of the initially filled sites (Fig~\ref{fig:SYK}). We compare exact dynamics to the results using mean-field and fTWA: while mean-field fails to capture the dynamics, the fTWA does a very accurate job. 
\begin{figure}[h]
  \centering
  \includegraphics[width=.8\linewidth]{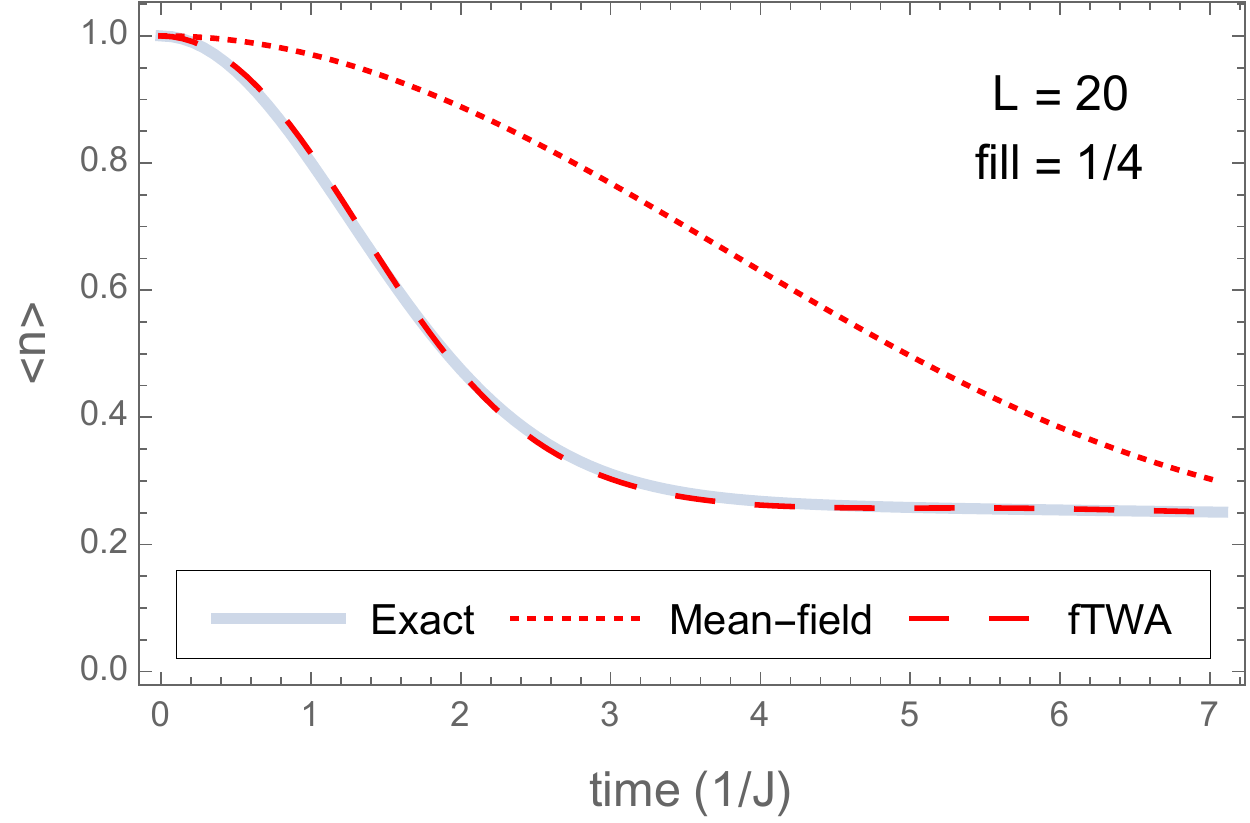}
  \caption{Thermalization dynamics for the SYK model with $L=20$ sites, with 5 sites initially filled in a product state. We look at the dynamics of the average occupancy of the initially filled sites. At long times, this approaches the average density, as all sites eventually appear the same regardless of initial conditions. However, the dynamics of the thermalization is not trivial. We plot the exact result, and compare to the mean-field result and fTWA. The mean-field is the dynamics calculated using a single classical trajectory, and can be thought of as the full classical limit. We see that this result deviates greatly from the exact result, while the fTWA, which we get by including quantum fluctuations in the initial condtions, is very close to the exact result.}
  \label{fig:SYK}
\end{figure}

\paragraph*{Conclusion}
In conclusion we presented a semiclassical phase-space method to simulate dynamics of interacting fermions, by constructing a phase-space out of all fermionic bilinears. In particular, we demonstrated that dynamics of fermions can be efficiently simulated through the classical Hamiltonian equations of motion of non-local string variables representing fermion bilinears. We applied these methods to nontrvial interacting setups, which are far from any mean-field regime or from the regime of validity of a quasi-particle description. The method moreover has the virtue of working in any dimension. This opens up the possibility of studying dynamics of strongly correlated fermions in condensed matter systems, quantum chemistry, and high-energy physics. Outstanding future problems include combining fTWA with efficient molecular dynamics type methods to allow accurate simulation of classical dynamics in macroscopic complex systems, and with ideas of Ref.~\cite{Davidson2015} by increasing the number of phase-space variables and controlling the error of the semiclassical approximation.

\paragraph*{Methods}
Before discussing fermions let us briefly review the phase-space
representation of quantum dynamics of interacting bosonic systems. The Wigner-Weyl correspondence provides a unique map between any function of
boson operators $\hat a$ and $\hat a^\dagger$ and a
function over the classical phase space of (complex) canonical
variables $\alpha$ and $\alpha^*$, called the Weyl
symbol of the operator:
\begin{align}
  \Omega_W(\vec \alpha) = \int {d \vec \eta d\vec \eta^\ast\over 2}
\langle \vec \alpha - \vec \eta/2 | \hat \Omega | \vec \alpha + \vec
\eta/2 \rangle \mathrm e^{(\vec\eta^* \vec\alpha - \vec \eta \vec
  \alpha^*)/2}.
\end{align}
The Weyl symbol of
the density matrix is known as the Wigner function~\cite{Polkovnikov2010a,Hillery1984}, which acts as a (quasi)-probability
distribution on phase space. Any quantum dynamical problem thus corresponds to a problem of propagating the Wigner function in time; its dynamics are determined by the phase-space analog of the commutator, called the Moyal bracket:
\begin{align}
  i \frac{d}{dt}W(\vec \alpha) &= \left\{H_W(\vec\alpha),W(\vec\alpha)\right\}_M \nonumber \\
  &= 2 H_W(\vec\alpha) \sinh\left(\Lambda_c/2 \right)W(\vec\alpha),
\end{align}
where $\Lambda_c$ is the coherent state Poisson bracket,
\begin{align}
  \Lambda_c=\sum_j
\frac{\overleftarrow{\partial}}{\partial\alpha_j} \frac{\overrightarrow{\partial}}{\partial\alpha_j^\ast}-
\frac{\overleftarrow{\partial}}{\partial\alpha_j^\ast} \frac{\overrightarrow{\partial}}{\partial\alpha_j}.
\end{align}

In general, time evolution of the Wigner function is hard to handle as the action for its propagator contains highly non-linear terms. However, near the classical limit one can use the so-called truncated Wigner approximation
(TWA)~\cite{Blakie2008,Polkovnikov2010a}, where we expand the Moyal bracket to first order:
\begin{align}
  i \frac{d}{dt}W(\vec \alpha) \approx  H_W(\vec\alpha) \Lambda_c W(\vec\alpha).
\end{align}
We recover the Liouville equation, and thus the Wigner
function, like a classical probability distribution, is conserved along
classical trajectories. The expectation value of any operator can
be straightforwardly computed at any moment in time:
 \begin{align}
\langle \hat
\Omega(t)\rangle\approx\int d\vec\alpha_0 d\vec\alpha_0^\ast
W(\vec\alpha_0,\vec\alpha_0^\ast)
\Omega_W(\vec\alpha_{cl}(t),\vec\alpha ^\ast_{cl}(t)),   
 \end{align}
where
$W(\vec\alpha_0,\vec\alpha_0^\ast)$ is the Wigner function
representing the initial state of the system,
$\Omega_W(\vec\alpha,\vec\alpha^\ast)$ is the Weyl symbol of the
operator $\hat \Omega$, and the classical paths are found by solving
Hamilton's equations,
\begin{align}
  \label{eq:cl}
  i \dot \alpha_{cl}=\frac{\partial H_W}{\partial \alpha_{cl}^\ast}.
\end{align}
In the bosonic case, TWA becomes more accurate as the number of particles increases. It is
always exact when the Hamiltonian is non-interacting, i.e. quadratic
in boson operators. For a more detailed discussion we refer to refs.~\cite{Blakie2008,Polkovnikov2010a}.

To construct the Wigner-Weyl correspondence for fermions we exploit the fact that the dynamical symmetry group of a physical fermionic system is comprised of bilinears of fermions. In fact, if we denote the fermionic bilinears by
\begin{gather}
  \label{eq:6}
  \hat E_{\alpha\beta} = \hat c_{\alpha} \hat c_{\beta}\\
  \hat E^{\alpha\beta} = \hat c^{\dagger}_{\alpha} \hat c^{\dagger}_{\beta}=-\hat E_{\alpha\beta}^\dagger\\
  \hat E^{\alpha}_\beta = \frac{1}{2}(\hat c^{\dagger}_\alpha \hat c_\beta - \hat c_\beta \hat c^{\dagger}_\alpha)=(\hat E^{\beta}_\alpha)^\dagger,
\end{gather}
and consider a system of $N$ fermionic modes, then these bilinears obey the Lie algebra of $so(2N)$~\cite{Fukutome1981}:
\begin{subequations}
\label{liealg}
\begin{align}
  [\hat E^\alpha_\beta,\hat E^\mu_\nu]_-&=\delta_{\beta\mu}\hat E^\alpha_\nu-\delta_{\alpha \nu}\hat E^\mu_\beta,\\
  [\hat E^\alpha_\beta,\hat E_{\mu \nu}]_-&=\delta_{\alpha \nu}\hat E_{\beta\mu}-\delta_{\alpha \mu}\hat E_{\beta \nu},\\
  [\hat E^{\alpha \beta},\hat E_{\mu \nu}]_-&=\delta_{\alpha \nu}\hat E^\beta_\mu+\delta_{\beta\mu}\hat E^\alpha_\delta -\delta_{\alpha \mu}\hat E^\beta_\nu-\delta_{\beta \nu}\hat E^\alpha_\mu,\\
  [\hat E_{\alpha \beta},\hat E_{\mu \nu}]_-&=0,\quad   [\hat E^{\alpha \beta},\hat E^{\mu \nu}]_-=0.
\end{align}
\end{subequations}
One can furthermore avoid working with rather complicated $so(2N)-$coherent states by first applying a Jordan-Schwinger map after which one can simply use ordinary bosonic coherent states to derive the TWA. Indeed, for any set of operators $\{\hat{X}_\alpha \}$ which satisfy commuation relations of some algebra $[\hat X_\alpha, \hat X_\beta]=i f_{\alpha\beta\gamma} \hat X_{\gamma}$, where $f_{\alpha\beta\gamma}$ are the structure constants, one can construct a Jordan map from the operator $\hat{X}_\alpha$ to the space of bosonic operators $\hat a_j$:
\begin{equation}
\hat{X}_\alpha \mapsto \hat a^\dagger_i[X_\alpha]^{ij} \hat a_j,
\end{equation}
where $[X_\alpha]^{ij}$ is the matrix representation of $\hat{X}_\alpha$. Clearly  both operators satisfy the same commutation relations.  As demonstrated in \cite{Davidson2015}, Wigner-Weyl quantization of the bosons and consequative truncated Wigner approximation results in a classical Poisson bracket for the original group operators of
\begin{gather}
\left\lbrace X_{\alpha}^\text{cl}, \mathcal{H} \right\rbrace_P=f_{\alpha\beta\gamma} {\partial \mathcal{H}\over \partial X^\text{cl}_{\beta}} X^\text{cl}_{\gamma},
\label{eq:TWAeqmot1}
\end{gather}
where $X_{\alpha}^\text{cl} = \left(a^\dagger_iX_\alpha^{ij}a_j\right)_W$ is the standard bosonic Weyl symbol of the bosonic bilinear form of the variable.

Specficially for fermions, we introduce the notation
 \begin{gather}
    \hat E^\alpha_\beta \mapsto \rho_{\alpha\beta},\\
  \hat E_{\alpha\beta}\mapsto \tau_{\alpha\beta} ,\quad     \hat E^{\alpha\beta} \mapsto -\tau_{\alpha\beta}^\ast,
\end{gather}
and the Poisson bracket has the form
\begin{gather}
  i\left\lbrace\rho_{\alpha\beta}, \mathcal{H} \right\rbrace_{fP} = \frac{\partial \mathcal{H}}{\partial \rho_{\beta \gamma}} \rho_{\alpha\gamma}-\frac{\partial \mathcal{H}}{\partial \rho_{\gamma\alpha}} \rho_{\gamma\beta}
+\frac{\partial \mathcal{H}}{\partial \tau_{\gamma\alpha}} \tau_{\beta \gamma}-\frac{\partial \mathcal{H}}{\partial \tau_{\alpha\gamma}} \tau_{\beta \gamma}
+\frac{\partial \mathcal{H}}{\partial \tau^*_{\beta\gamma}} \tau^*_{\alpha \gamma}-\frac{\partial \mathcal{H}}{\partial \tau^*_{\gamma\beta}} \tau^*_{\alpha \gamma}\\
  i\left\lbrace\tau_{\alpha\beta} , \mathcal{H} \right\rbrace_{fP}= \frac{\partial \mathcal{H}}{\partial \rho_{\alpha \gamma}} \tau_{\gamma\beta}-\frac{\partial \mathcal{H}}{\partial \rho_{\beta\gamma}} \tau_{\gamma\alpha}
+\frac{\partial \mathcal{H}}{\partial \tau^*_{\gamma\alpha}} \rho_{\gamma\beta}+\frac{\partial \mathcal{H}}{\partial \tau^*_{\beta\gamma}} \rho_{\gamma\alpha}
-\frac{\partial \mathcal{H}}{\partial \tau^*_{\alpha\gamma}} \rho_{\gamma\beta}-\frac{\partial \mathcal{H}}{\partial \tau^*_{\gamma\beta}} \rho_{\gamma\alpha}.
\label{eq:fP}
\end{gather}
It is easy to check that these formal rules agree with a more intuitive pictorial representation of the Poisson brackets (see Eq.~\eqref{eq:poisson_fermion}).

Within the accuracy of TWA, instead of the exact Wigner function it suffices to use a Gaussian distribution, with mean and variance fixed by the initial conditions of the initial quantum state. Numerically this has a huge advantage, as propagating the Wigner function now amounts to solving equation of motion \eqref{eq:TWAeqmot} for random initial conditions $\rho_{\alpha\beta}$ and $\tau_{\alpha\beta}$, drawn from a normal distribution. When we fix the variance of the classical fermionic variables, we have to symmetrize their quantum analogs, as the classical variables commute:
\begin{gather}
  \braket{\hat E^\alpha_\beta} = \int d \rho d \tau \mathcal{W}(\rho,\tau)\rho_{\alpha\beta},\quad \braket{\hat E_{\alpha\beta}} = \int d \rho d \tau \mathcal{W}(\rho,\tau)\tau_{\alpha\beta},\\
\braket{(\hat E^\alpha_\beta)^\dagger \hat E^\mu_\nu+ \hat E^\mu_\nu(\hat E^\alpha_\beta)^\dagger}/2 = \int d \rho d \tau \mathcal{W}(\rho,\tau)\rho_{\alpha\beta}^*\rho_{\mu\nu},\\
\braket{(\hat E_{\alpha\beta})^\dagger \hat E_{\mu\nu}+ \hat E_{\mu\nu}(\hat E_{\alpha\beta})^\dagger}/2 = \int d \rho d \tau \mathcal{W}(\rho,\tau)\tau_{\alpha\beta}^*\tau_{\mu\nu}.
\end{gather}
Any time-dependent expectation value can be calculated by averaging over these initial conditions: 
\begin{align}
\langle \hat
O(t)\rangle\approx\int d \rho d \tau
\mathcal{W}(\rho,\tau)
\mathcal{O}(\rho_{cl}(t),\tau_{cl}(t)).
\label{eq:TWAexp} 
 \end{align}

For the interaction terms of the Hamiltonians, with 4 fermion terms, there are multiple ways to apply the Jordan-Schwinger mapping in terms of bifermion operators, and these will give us different Weyl Hamiltonians. For example, consider the interaction term
\begin{subequations}
\begin{align}
  \hat V_{ijkl} &= \hat c^\dagger_i\hat c^\dagger_j \hat c_k\hat c_l\\
&= \hat E^{ij} \hat E_{kl} \label{eq:tau}\\
&= -\hat E^i_k \hat E^j_l + \delta_{jk}\hat E^i_l \label{eq:rho}
\end{align}
\end{subequations}
While both forms \eqref{eq:tau} and \eqref{eq:rho} correspond to the same operator, they map to different Weyl symbols in the classical language. To find the Weyl symbol of quartic operators, it is useful to introduce Bopp operators~\cite{Polkovnikov2010a}, which allow us to find the Weyl symbol of composite operators. For the fermionic string variables, they are
\begin{gather}
  \label{eq:1}
  \hat E^\alpha_\beta \mapsto \rho_{\alpha\beta} 
+ \frac{1}{2} \left(\rho_{\alpha\gamma} \frac{\partial}{\partial \rho_{\beta\gamma}} - \rho_{\gamma\beta} \frac{\partial}{\partial \rho_{\gamma\alpha}}
+ \tau_{\beta\gamma} \left(\frac{\partial}{\partial \tau_{\gamma \alpha}} - \frac{\partial}{\partial \tau_{\alpha \gamma}} \right)  + \tau^*_{\alpha \gamma} \left(\frac{\partial}{\partial \tau^*_{\beta \gamma}} - \frac{\partial}{\partial \tau^*_{\gamma \beta}}\right)\right)\\
 \hat E_{\alpha\beta}\mapsto \tau_{\alpha\beta} + \frac{1}{2}\left( \tau_{\gamma \beta} \frac{\partial}{\partial \rho_{\gamma\alpha}} - \tau_{\alpha\gamma} \frac{\partial}{\partial \rho_{\beta \gamma}}
 + \rho_{\gamma \beta} \left(\frac{\partial}{\partial \tau^*_{\gamma\alpha}} - \frac{\partial}{\partial \tau^*_{\alpha\gamma}} \right)  + \rho_{\gamma\alpha} \left(\frac{\partial}{\partial \tau^*_{\beta\gamma}} - \frac{\partial}{\partial \tau^*_{\gamma\beta}}\right)\right),
\end{gather}
Just like for ordinary Weyl symbol~\cite{Polkovnikov2010a} one can define a left acting version of them by changing the sign in front of the bracket and having all the derivatives act to the left. They allow us to easily find the Weyl symbol of composite operators and naturally lead to the fermionic Poisson bracket (\eqref{eq:fP}), while they also make clear that symmetric ordering of operators maps directly to classical variables.

Using the Bopp operators, we can map form \eqref{eq:tau} of the interaction to classical $\tau$ varables,
\begin{gather}
  \label{eq:3}
  \left(\hat E^{ij} \hat E_{kl}\right)_W = \tau^*_{ji} \tau_{kl} + \frac{1}{2} \rho_{jk} \delta_{il} - \frac{1}{2} \rho_{jl} \delta_{ik} + \frac{1}{2} \rho_{il} \delta_{kj} - \frac{1}{2} \rho_{ik} \delta_{jl}
\end{gather}
while form \eqref{eq:rho} of the interaction maps to $\rho$ variables:
\begin{align}
  \left(-\hat E^i_k \hat E^j_l + \delta_{jk}\hat E^i_l\right)_W =-\left(\rho_{ik} + \delta_{ik}/2 \right)\left(\rho_{jl} + \delta_{jl}/2\right)
+ \frac{1}{2} \delta_{jk} \rho_{il} + \frac{1}{2} \delta_{il} \rho_{jk} + \delta_{jk}\delta_{il}/2.
\end{align}
The approximate dynamics of fTWA will be affected on the choice of variables, and depending on the problem, a different variable choice may lead to a better approximation. 

In the SYK example shown in the text (Fig~\ref{fig:SYK}), we choose to use the Hamiltonian with $\tau$ variables as the interaction,
\begin{align}
  \mathcal{H}_\tau &= \frac{1}{(2N)^{3/2}}\sum_{ijkl} J_{ij;kl} \left( \tau^*_{ji} \tau_{kl} + \frac{1}{2} \rho_{jk} \delta_{il} - \frac{1}{2} \rho_{jl} \delta_{ik} + \frac{1}{2} \rho_{il} \delta_{kj} - \frac{1}{2} \rho_{ik} \delta_{jl} \right).
\end{align}
As we have seen, we could have alternatively mapped to an interaction term in terms of $\rho$ variables,
\begin{align}
  \mathcal{H}_\rho &=\frac{1}{(2N)^{3/2}}\sum_{ijkl} J_{ij;kl} \bigg(-\left(\rho_{ik} + \delta_{ik}/2 \right)\left(\rho_{jl} + \delta_{jl}/2\right)\\
 &\qquad \qquad \qquad  + \frac{1}{2} \delta_{jk} \rho_{il} + \frac{1}{2} \delta_{il} \rho_{jk} + \delta_{jk}\delta_{il}/2\bigg),
\end{align}
or we could also use both possible options, and use the Hamiltonian
\begin{gather}
  \mathcal{H}_+ = (\mathcal{H}_\rho + \mathcal{H}_\tau)/2.
\end{gather}
In Fig.~\ref{fig:SYK_all} we have simulated dynamics for the same model using these three possible choices of Hamiltonian. Clearly using the superconducting $\tau$ variables in the interaction is the correct choice; it is an open question as to why this is the case. Likewise in the expansion example, where we do not expect superconducting variables to play any role, we found that the representation of the interaction through $\rho$ variables leads to a much better approximation.
\begin{figure}[h]
  \centering
  \includegraphics{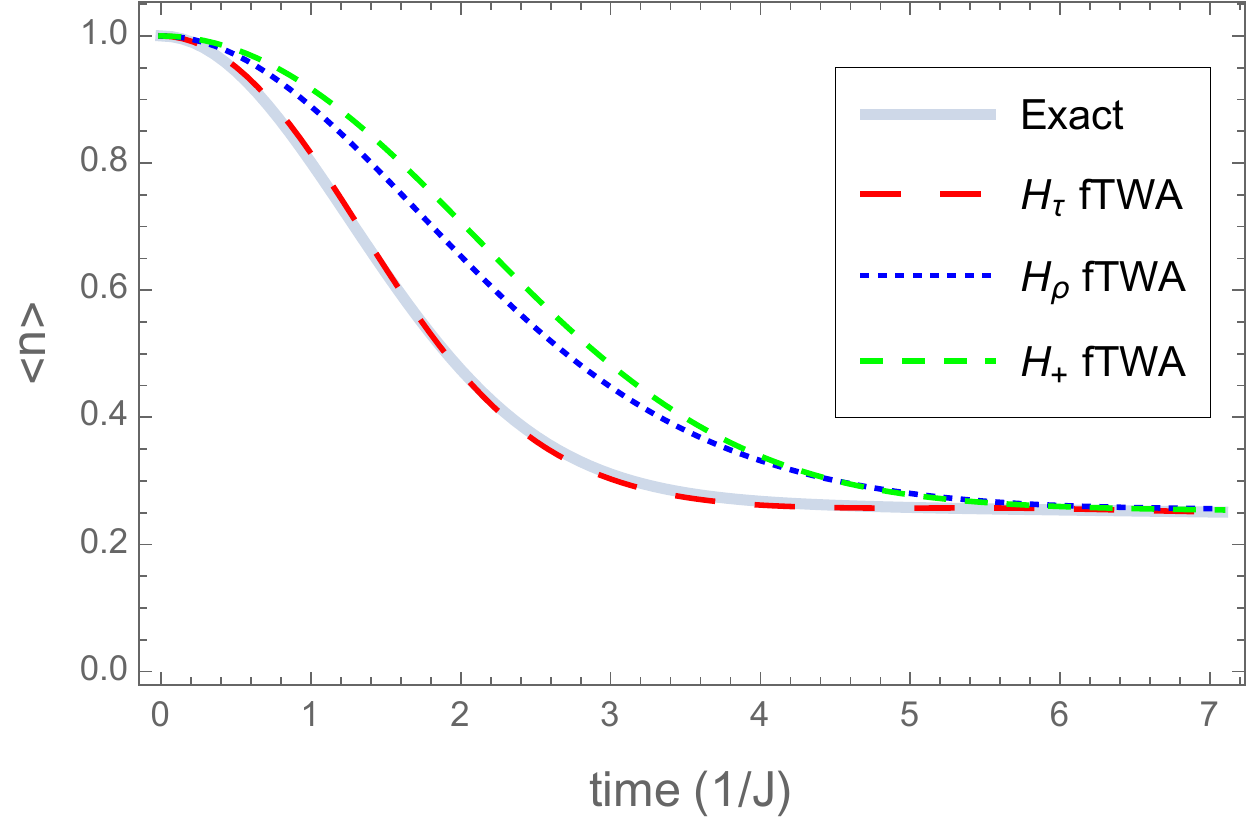}
  \caption{Exact dynamics for the SYK model compared to fTWA dynamics using the three possible forms of the classical Hamiltonian described in the text. The model is the same as in the main text, with $L=20$ sites and starting with 5 sites filled in a product state. We measure the average occupancy of the initially filled sites. The Hamiltonian containing only an interaction between the superconducting $\tau$ variables appears to be the correct semi-classical limit}
  \label{fig:SYK_all}
\end{figure}


\acknowledgements
We thank Valentin Kasper for collaboration in the early stages of this work. 
D.S. acknowledges support of the FWO as post-doctoral fellow of the Research Foundation - Flanders. A.P. and S. D. were supported by AFOSR FA9550- 13-1-0039, NSF DMR-1506340 and ARO W911NF1410540.


\bibliography{FermionTWAPaperFinal}

\pagebreak
\widetext
\begin{center}
\textbf{\large Supplementary Information:\\Semiclassical Approach to Dynamics of Interacting Fermions}
\end{center}
\begin{center}
S. Davidson, D. Sels, A. Polkovnikov \\
%\emph{Department of Physics, Boston University, MA 02215, USA}
\end{center}
\setcounter{equation}{0}
\setcounter{figure}{0}
\setcounter{table}{0}
\setcounter{page}{1}
\renewcommand{\theequation}{S\arabic{equation}}
\renewcommand{\thefigure}{S\arabic{figure}}
\renewcommand{\bibnumfmt}[1]{[S#1]}

\date{\today}
\maketitle
In the main text we discussed two models, the Hubbard model and SYK model; here we present results on a third model, combining normal bosonic TWA with fTWA.
\section*{Two Channel model}
We consider the Hamiltonian representing fermions resonantly coupled to a bosonic molecular state: 
\begin{equation}
\hat{H} =\sum_{i}\mu_{B}b_{i}^{\dagger}b_{i}-J\sum_{\sigma<ij>}\left(  c_{\sigma i}^{\dagger
}c_{\sigma j}+\text{h.c.}\right) +g\sum_{i} \left(b_{i}c_{\uparrow i}^{\dagger}c_{\downarrow i}^{\dagger}+\text{h.c.}\right),
\label{eq:ham}
\end{equation}
with bosons satisfying $[b_i,b^\dagger_j]_-=\delta_{ij}$ and fermions satisfying $[c_{\sigma i},c^\dagger_{\sigma^\prime j}]_+=\delta_{\sigma \sigma^\prime}\delta_{ij}$. This Hamiltonian describes very well interacting fermions near the Feshbach resonance (see e.g. Ref.~\cite{Andreev2004}). For large positive (negative) chemical potential the molecular state can be integrated out (provided that it is not populated) and this Hamiltonian reduces to the attractive (repulsive) Hubbard model. This model also describes the BCS-BEC crossover as one gradually tunes $\mu_B$ from a positive to a negative value. Close to mean-field regimes, where the bosonic field condenses, this model is amenable to various analytic treatments~\cite{Andreev2004, Altman2005}, but far from the mean-field limit and far from equilibrium it essentially cannot be simulated with existing numerical or analytical tools.

%Note that for this model we can use a subgroup of $so(2N)$: we only include $\hat E^\alpha_\beta$ where $\alpha$ and $\beta$ correspond to fermions with the same spin, and $\hat E_{\alpha\beta}$ where $\alpha$ and $\beta$ correspond to fermions with different spin. The corresponding Weyl Hamiltonian is
%\begin{multline}
%  H =\sum_{i}\mu_{B}\beta_{i}^*\beta_{i}\\
%-J\sum_{\sigma<ij>}\left(\rho_{\sigma ij}+\rho_{\sigma ij}^*\right)+g\sum_{i} \left(\beta_{i}\tau_{ii}^*+\beta_{i}^*\tau_{ii}\right).
%\end{multline}

\begin{figure}
  \centering
  \includegraphics[width=0.45\textwidth]{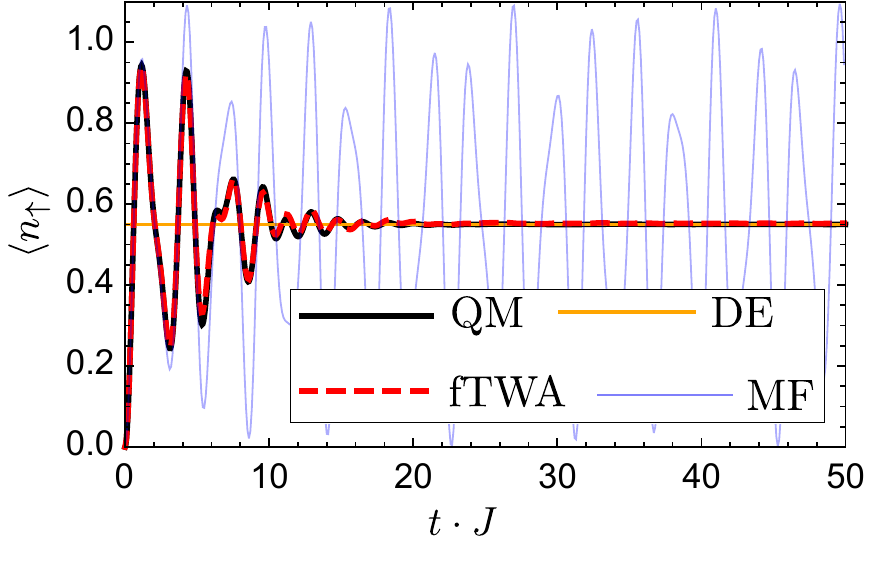}
  \caption{Simulation of the quench dynamics for the two-site Hamiltonian \eqref{eq:ham} with $\mu_B = 1$ and $g=1/3$. The initial state has no fermions and a coherent state of bosons on each site with average number of nine bosons per site. We measure the average number of spin up fermions. The black line labelled (QM) represents the exact quantum evolution. The mean-field (MF) result is accurate at short times, but the fTWA result continues to be very accurate at all times, reproducing the final steady state, which is that predicted by the diagonal ensemble (DE).}
  \label{fig:2site}
\end{figure}

We express this Hamiltonian in terms of the $\rho_{\alpha\beta}$ and $\tau_{\alpha\beta}$ where $\alpha$ and $\beta$ label the site component and the spin index. As a first demonstration, we look at a system of two sites. In Fig. \ref{fig:2site}, we compare exact quantum dynamics to those using classical equations of motion. The initial quantum state is a vacuum for fermions and a coherent state for bosons on each of the two sites with a mean number of bosons of $N_i=9$ per site. The Wigner function for the bosons is thus a product of two Gaussians, $\mathcal{W}(b_i,b_i^\ast)=2\exp[-2|b_i-\sqrt{N_i}|^2]$ (see e.g. Ref.~\cite{Polkovnikov2010a}), and the Wigner function of the bilinears is also approximated by the products of Gaussians. We deliberately choose rather large initial boson occupation number per site to be in the regime where the method is expected to be nearly exact. We quench to $\mu_B = 1$ and $g=1/3$, and show the corresponding rise in the average number of fermions.

For the classical dynamics, we show both mean-field (MF) initial conditions (where we calculate only one classical path, with each classical variable determined by the average of the corresponding quantum operator) along with the full fTWA (where we integrate over many different initial condition determined by the Wigner function). Even though in this example one can naively expect the MF approximation to be rather accurate it is clear that the fTWA gives far better results, because we include the correct initial correlations. We even predict the saturation to the correct final steady state, which agrees with the quantum diagonal ensemble. So in this case the method does not have typical short time limitations~\cite{Polkovnikov2010a}. Qualitatively this can be understood from the fact that at long times the system goes to a highly excited (and highly entangled) classical state, where quantum fluctuations are small.

\begin{figure*}
  \centering
  \includegraphics[width=\linewidth]{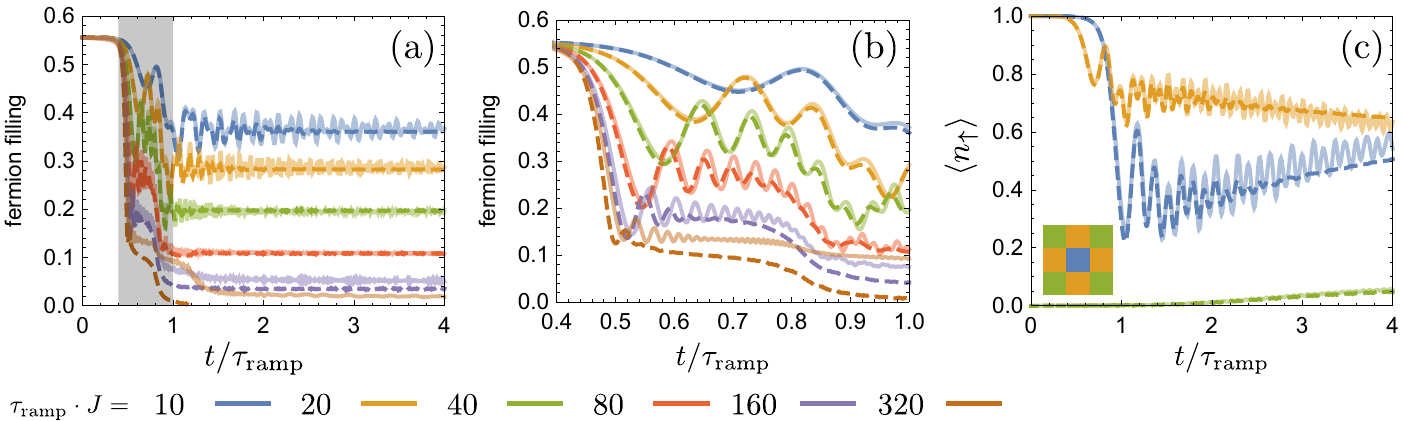}
  \caption{Exact quantum dynamics (solid) compared with fTWA (dashed) for a $3\times 3$ lattice. Panels (a) and (b): average fermionic filling fraction, with $\tau_\text{ramp} =10,20,40,80,160,320$ in units of the coupling $J$. Panel (c):  the average fermion number for each mode for $\tau_\text{ramp} =10$ (inset matches colors to location on the Fermi-surface). }
  \label{fig:3b3}
\end{figure*}

To further demonstrate the method on a larger 2D system, we compare fTWA results with the exact case on a nine site $3\times 3$ lattice with periodic boundary conditions. Here we start in the ground state of the Hamiltonian with $\mu_B=0$ and $g=0$, with no bosons and a Fock state of fermions in momentum space filling up the five lowest energy modes. Note that this is the worst regime for the bosonic TWA, as TWA approximations generally get better with higher particle number, i.e. the TWA is a $1/N$ expansion. Unlike the previous two-site example, there is no obvious small parameter here.  We then ramp the chemical potential and the coupling, with $\mu_B(t)=-10(1-e^{-(t/\tau_\text{ramp})^2})$ and $g(t)=1-e^{-(t/\tau_\text{ramp})^2}$.
By controlling the ramp time, we control the final diagonal entropy of the state. This allows us to investigate the accuracy of the fTWA. 
 For large enough $\tau_\text{ramp}$, the process would be adiabatic and we would end up with nearly all fermions converted to bosons, as we are moving deep into the BEC regime. In Fig. \ref{fig:3b3} we show results comparing the average fermion filling fraction, with $\tau_\text{ramp}$ ranging from 10 to 320 in units of the coupling $J$. As the time period extends, the approximation begins to break down as generally expected for TWA~\cite{Polkovnikov2010a}. Physically this corresponds to the fact that the ground state has stronger quantum fluctuations. At very slow ramps the approximation even yields unphysical negative occupation numbers. At the same time the method gives very accurate predictions at intermediate ramp rates where the majority of the fermions are converted to bosons such that short time perturbative expansions completely fail. So while with fTWA we cannot correctly predict the ground state of this model, we are able to accurately describe both transient dynamics and the steady-state in other highly non-trivial strongly-correlated regimes.

\begin{figure}
  \includegraphics[width=0.5\linewidth]{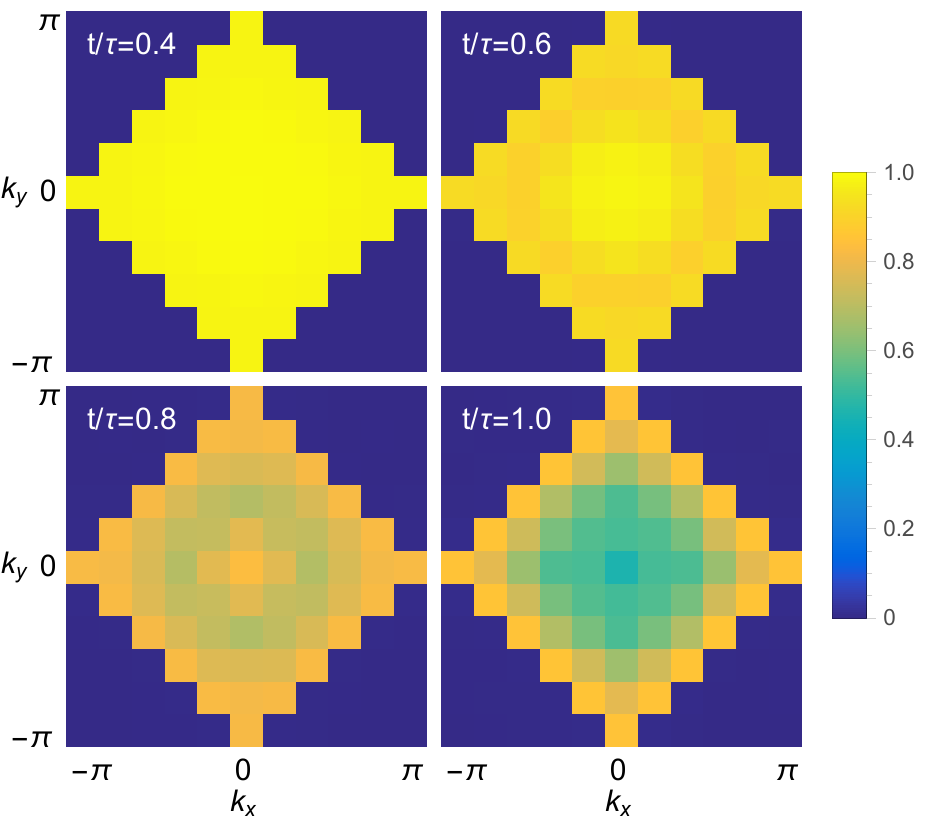}
  \caption{Results for the Fermi-surface of a $10\times 10$ system with $\tau_\text{ramp}=10$, at times $t/\tau_\text{ramp}=0.4,0.6,0.8,1.0$. Fermions near the Fermi-surface start converting to bosons first, but in the end the lowest energy fermions end up being most likely to convert, resulting in an inverted non-equilibrium final fermion distribution: the highest energy fermions are too fast to combine and convert into molecules.}
  \label{fig:10b10}
\end{figure}

Finally, we demonstrate the power of fTWA for a system that cannot be simulated exactly: in Fig. \ref{fig:10b10}, we extend the previous example to a larger $10\times 10$ system and analyze the time evolution of the Fermi-surface. In this case, we use $\tau_\text{ramp}=10$, where fTWA is expected to be nearly exact, and show snapshots of the Fermi-surface when it is most volatile, from $t/\tau_\text{ramp}=0.4-1$. We see that the fermions near the Fermi-surface start converting to bosons first, but then quickly the lowest energy fermions end up being most likely to convert, resulting in an inverted non-equilibrium final fermion distribution. Physically this unusual final state indicates that the highest energy fermions are too fast to combine and convert into molecules. This is consistent with the exact results we see in the $3\times 3$ lattice (Fig. \ref{fig:3b3}(c)). 

%In the Supp. Material~\cite{supplement} we further demonstrate the power of fTWA by simulating quench dynamics in disordered MBL systems, mimicking a recent experiment of Ref.~\cite{Fischer2015} and extending simulations to a two-dimensional regime.

% \acknowledgements
% D.S. acknowledges support of the FWO as post-doctoral fellow of the Research Foundation - Flanders. A.P. and S. D. were supported by AFOSR FA9550- 13-1-0039, NSF DMR-1506340 and ARO W911NF1410540.


\end{document}